\documentclass[12pt]{article}
\usepackage{amsmath}
\usepackage{latexsym}
\usepackage{amssymb}
\newtheorem{defi}{Definition}
\newtheorem{lem}[defi]{Lemma}
\newtheorem{thm}[defi]{Theorem}
\makeatletter
\newenvironment{pf}
{\noindent{\bf Proof}\quad}{\leavevmode\hfill$\Box$\par\@endpetrue}
\makeatother

\def\Tr{\mathop{\rm Tr}\nolimits}

\def\real{\mathbb{R}}

\def\SU{\mathop{\rm SU}}
\def\sgn{\mathop{\rm sgn}}
\def\id{\mathop{\rm id}}
\def\Label#1{\label{#1}\ [\ #1\ ]\ }
\def\Label{\label}
\begin{document}
\bibliographystyle{unsrt}
\newpage
%\twocolumn[
\begin{center}
{\Large\bf
Variable length universal entanglement
concentration\par by local operations and\par
its application to teleportation and dense coding
\par} \vskip 1.5em 
Masahito Hayashi \par 
Laboratory for Mathematical Neuroscience, 
Brain Science Institute, RIKEN\par
2-1 Hirosawa, Wako, Saitama, 351-0198, Japan\par
Keiji Matsumoto\par
Quantum Computation and Information Project, ERATO, JST\par
5-28-3, Hongo, Bunkyo-ku, Tokyo, 113-0033, Japan
\end{center}
%]
\begin{abstract}
Using invariance of 
the $n$-th tensored state w.r.t. the $n$-th symmetric group, 
we propose a 'variable length' 
universal entanglement concentration without classical 
communication.
Like variable length data compression,
arbitrary unknown states are
concentrated into perfect Bell states and  not approximate
Bell states and the number of Bell states obtained is 
equal to the optimal rate asymptotically with the probability 1. 
One of the point of our scheme is that we need
no classical communication at all. 
Using this method, we can 
construct a universal teleportation 
and a universal dense coding.
\end{abstract}
\section{Introduction}\Label{s1}
In quantum systems, we can perform some 
information processes which do not appear in classical systems.
For example, quantum teleportation, dense coding etc.
For them it is necessary to share an entangled state between 
two systems.
If the entangled state is the perfect Bell state,
its analysis is very easy.
Otherwise, it is not easy \cite{B2,B1}.

We can produce perfect Bell states from arbitrary 
entangled states by local operations and classical communications (LOCC)
and call such an operation {\it an entanglement concentration}.
As is proved by Bennett et al\cite{Ben},
when we share the $n$-tensor product state 
$|\phi \rangle \langle \phi|^{\otimes n}$
on the total tensor product system
${\cal H}_A^{\otimes n} \otimes {\cal H}_B^{\otimes n}$,
we can produce, by local operations, 
$n H(\rho_A)$-qubit perfect Bell states 
asymptotically with the probability $1$,
where $\rho_A:= \Tr _B |\phi \rangle \langle \phi|$ and
$H(\rho_A)$ is the entropy $- \Tr \rho_A \log \rho_A$.

In this paper, we propose 
{\it a 'variable length' universal  entanglement 
concentration} without any classical communication.
Like variable length data compression,
arbitrary unknown states are
concentrated into perfect Bell states and  not approximate
Bell states, and the number of Bell states obtained is
equal to $n H(\rho_A)$ asymptotically with the probability 1. 
One of the point of our scheme is that we  need
no classical communication at all. 

In \S \ref{s2}, we propose {\it a variable length group-invariant entanglement
concentration} consisting of local operations
when the entanglement pure state is invariant w.r.t. the 
tensor representation on ${\cal H}_A \otimes {\cal H}_B$ 
of a group $G$, where ${\cal H}_A$ and ${\cal H}_B$
are equivalent with each other w.r.t. a representation space of $G$.
In this method, the final state is always the perfect Bell state
and the size is probabilistic.
In \S \ref{s3} using invariance of 
the $n$-th tensored state w.r.t. the $n$-th symmetric group, 
we construct a variable length universal entanglement concentration
(simplified to a universal entanglement concentration),
in which, we can,
independently of $\rho_A$, 
produce no less than $n H(\rho_A)$-qubit perfect Bell states 
asymptotically with the probability $1$.
As another method,
we can perform an entanglement concentration 
after the state estimation on $\epsilon n$ systems.
But, if we perform entanglement concentration which depends on the
estimated state, 
the final state is not necessarily the perfect Bell state
because the estimated state does not exactly coincide
with the true state.
As is proved in \S \ref{s3.1}, our universal concentration 
achieves the optimal failure exponent
among universal concentrations which achieve the optimal rate 
$n H(\rho_A)$ for any state 
asymptotically with the probability $1$.

In the quantum teleportation, we can send a quantum state with LOCC.
In such a setting we maximize the number 
of teleported qubits only with LOCC.
As is discussed in \S \ref{s4} to share $R$-qubit perfect Bell state
is equivalent with
to send $R$-qubit of perfect Bell states only with LOCC.
Therefore, we can perform $n H(\rho_A)$ qubits quantum teleportation,
under the assumption that we share the $n$-tensor product state 
$|\phi \rangle \langle \phi|^{\otimes n}$
on the total system ${\cal H}_A^{\otimes n} \otimes {\cal H}_B^{\otimes n}$.
Even if we do not know the density operator $\rho_A$,
using our universal entanglement concentration
we can perform $n H(\rho_A)$ qubits quantum teleportation
asymptotically with the probability $1$.
In the protocol, it is enough to send the minimum classical communications
of the size of $2 n H(\rho_A)$ bits.

If entangled states are shared,
we can send $R_1$ bits classical message
by sending only $R_2 ( \,< R_1)$ qubits.
This type information process is called (super) dense coding.
The number $R_1-R_2$ signifies the effect of entanglement.
Thus, in this setting we can regard the maximum of $R_1-R_2$ as
the capacity.
Our setting is different from the usual setting of the dense coding.
As is discussed in \S \ref{s5}, we can prove that the maximum of
$R_1-R_2$ is asymptotically equal to $n H(\rho)$.
Even if we do not know the density $\rho_A$,
using our universal entanglement concentration
we can send $2n H(\rho)$ bits of classical information
by sending only $n H(\rho)$ qubits.

As is pointed out by Keyl and Wener \cite{KW},
this group invariant method is applicable to 
the estimation of spectrum.
Concerning this topic, we will discuss another paper\cite{suitei}.
\section{Variable length group-invariant entanglement concentration}\label{s2}
For the preparation of our universal entanglement concentration,
we construct a entanglement concentration protocol 
under the group representation-invariance in a non-asymptotic
setting.
We call this protocol {\it a variable length group-invariant entanglement
concentration}
(simplified to {\it an invariant entanglement concentration}).
Let $f_A$ and $f_B$ be unitary representations of a group $G$
on finite dimensional spaces ${\cal H}_A$ and ${\cal H}_B$, 
which are equivalent with each other.
Assume that we share 
the pure state $| \phi \rangle \langle \phi |$
which is invariant under the tensor representation 
$f_A \otimes f_B$ 
on the total system ${\cal H}_A \otimes {\cal H}_B$,
i.e. 
$f_A(g) \otimes f_B(g) | \phi \rangle = | \phi \rangle,
\forall g \in G$.
\begin{lem}\Label{l1}
If $f_A$ and $f_B$ are irreducible,
the invariant vector
$\phi$
is given as
\begin{align*}
|\phi \rangle =
\frac{1}{\sqrt{d}}\sum_{j=1}^{d}
|e_{j,A} \rangle \otimes |e_{j,B} \rangle,
\end{align*}
where $\{e_{j,A}\}_{j=1}^{d}$ and $\{e_{j,B}\}_{j=1}^{d}$
are CONSs of ${\cal H}_A$ and ${\cal H}_B$ such that 
$f_{is}(e_{j,A})= e_{j,B}$, where
$f_{is}$ is the unique isomorphism map from ${\cal H}_{A}$ to ${\cal H}_{B}$,
w.r.t. the representation of $G$.
With ambiguity of constant factor,
the vector $| \phi \rangle$
is uniquely defined from the invariance of the representation of $G$.
\end{lem}
Then, we call the vector $\phi$ {\it the invariant perfect Bell state}
on ${\cal H}_A \otimes {\cal H}_B$.

\begin{pf}
Since ${\cal H}_A$ and ${\cal H}_B$ are equivalent w.r.t. the
representation space of $G$,
we can identify the
space ${\cal H}_A \otimes {\cal H}_B \cong {\cal H}_A \otimes {\cal H}_B^*$ with 
the set ${\cal L}({\cal H})$ of linear transforms on ${\cal H}_A$.
In this identification,
the representation of $G$ on ${\cal H}_A \otimes {\cal H}_B$ is
regarded as the adjoint representation on ${\cal L}({\cal H})$
because 
$f_A(g) f_B(g) |\phi_A\rangle \otimes |\phi_B\rangle 
\cong f_A(g) |\phi_A\rangle \otimes \langle \phi_B| f_B(g)^*
\cong f_A(g) |\phi_A\rangle \otimes \langle \phi_B| f_A(g)^{-1}, $\quad
$\forall |\phi_A\rangle \in {\cal H}_A, \forall |\phi_B\rangle \in {\cal H}_B$.
Therefore, using Schur's lemma, we can prove the desired assertion.
\end{pf}
Since the dimension of ${\cal H}_A$ is finite,
there exists a decomposition into irreducible representations
of ${\cal H}_A$ as follows:
\begin{align}
{\cal H}_A &=
\bigoplus_k \left( \bigoplus_{i=1}^{l_k} V_{k,i,A} \right) \Label{h1}\\
{\cal H}_B &=
\bigoplus_k \left( \bigoplus_{i=1}^{l_k} V_{k,i,B} \right) \Label{h1.1},
\end{align}
where $V_{k,i}$ and $V_{k,j}$ is equivalent w.r.t. the representation of $G$.
Therefore,
there are $l_k$ spaces equivalent with $V_{k,1}$
w.r.t. the representation of $G$.
Note that the decomposition 
is not unique, if there is a pair of equivalent 
subspaces.
Let $U_{k,A}$ and $U_{k,B}$ be the vector spaces 
$\langle e_{k,1,A}, \ldots, e_{k,{l_k},A} \rangle$
and $\langle e_{k,1,B}, \ldots, e_{k,{l_k},B} \rangle$.
and $V_{k,A}$ and $V_{k,B}$ be a vector space equivalent 
with $V_{k,i,A}$, $V_{k,i,B}$ w.r.t. the 
representation of $G$.
Then we have 
\begin{align}
{\cal H}_A &=\bigoplus_k U_{k,A} \otimes V_{k,A} \Label{h4} \\
{\cal H}_B &=\bigoplus_k U_{k,B} \otimes V_{k,B}.\Label{h4.1}
\end{align}
\begin{lem} From the invariance of $f_A \otimes f_B$,
we can choose the decomposition (\ref{h1}) and (\ref{h1.1})
satisfying that
\begin{align}
|\phi \rangle=
\sum_k \sum_{i=1}^{l_k}
\sqrt{s_{k,i} d_k } 
|\phi^P_{k,i}\rangle
\Label{h2}
\end{align}
where $d_k= \dim V_k$, and the vector 
$\phi^P_{k,i} \in V_{k,i,A} \otimes V_{k,i,B}$
is the invariant 
perfect Bell state on $V_{k,i,A} \otimes V_{k,i,B}$.
\end{lem}
\begin{pf}
Similarly to Lemma \ref{l1}, 
using Schur's lemma, we can prove the desired assertion.
\end{pf}
The constant factor $s_k$ satisfies that
\begin{align*}
\rho_A= \sum_{k}\sum_{i=1}^{l_k}
s_{k,i} V_{k,i,A} ,
\end{align*}
where we identify the subspace of ${\cal H}_A$ with 
its projection and $\rho_A := \Tr_B | \phi \rangle \langle \phi|$.
We cannot choose the decompositions
(\ref{h1}) and (\ref{h1.1}) satisfying (\ref{h2})
from the invariance of $f_A \otimes f_B$.
But, can uniquely construct the decompositions (\ref{h4}) and
(\ref{h4.1}) from the invariance of $f_A \otimes f_B$.

Let us construct the invariant entanglement concentration.
First, we perform the projection measurements
$\{U_{k,A} \otimes V_{k,A} \}_k $ and $\{U_{k,B} \otimes V_{k,B} \}_k$
on ${\cal H}_A$ and ${\cal H}_B$,
i.e. we perform the projection measurement
$\{U_{k_A,A} \otimes V_{k_A,A} \otimes U_{k_B,B} \otimes V_{k_B,B}
\}_{k_A,k_B}$
on the total system ${\cal H}_A \otimes {\cal H}_B$.
It follows from (\ref{h2}) that 
the event $k_A \neq k_B$ happens with the probability $0$
and the event $k_A = k_B= k$ happens with the probability 
$c_k:= d_k \sum_{i=1}^{l_k} s_{k,i}$.
If the measured value $k_A=k_B$ is $k$,
the state on $U_{k_A,A} \otimes V_{k_A,A} \otimes U_{k_B,B}
\otimes V_{k_B,B} $
is written by  
\begin{align*}
\frac{1}{\sqrt{c_k}} \sum_{i=1}^{l_k}
\sqrt{s_{k,i}d_k}
|\phi^P_{k,i} \rangle .
\end{align*}
Next, we take the partial trace on $U_{k,A} \otimes U_{k,B}$.
Then the final state is
the invariant perfect Bell state on 
$U_{k,A} \otimes U_{k,B}$,
whose size is $d_k= \dim V_k$.
Using this protocol, we can get the perfect Bell state with the $\dim V_k$
in the probability $c_k:= \dim V_k \sum_{i=1}^{l_k} s_{k,i}=
\Tr \rho_A U_k \otimes V_k$.

\section{Universal entanglement concentration}\label{s3}
It is well-known that the tensor product state
is invariant under the representation of $n$-th symmetric group.
Applying the invariant entanglement concentration to this case,
we can construct a universal entanglement concentration.
Let $d$ be the maximum of $\dim {\cal H}_A$ and $\dim {\cal H}_B$.
We add some vectors so that the relation 
$d = \dim {\cal H}_A = \dim {\cal H}_B$ holds.

We assume that 
the state on the tensored total system 
${\cal H}_A^{\otimes n} \otimes {\cal H}_B^{\otimes n}$
is written by $n$-tensored state $| \phi \rangle \langle \phi 
|^{\otimes n}$,
where $| \phi \rangle \langle \phi |$ is a pure state on 
the single total system ${\cal H}_A \otimes {\cal H}_B$.
Define the subscript ${\bf n}$ by 
\begin{align*}
{\bf n}: = (n_1, \ldots, n_d) , \quad \sum_{i=1}^d
n_i= n, n_{i} \ge n_{i+1}.
\end{align*}
The subscript ${\bf n}$ uniquely corresponds to 
the unitary irreducible representation $V_{{\bf n}}$ of the $n$-th symmetric group $S_n$ 
and the unitary irreducible representation $U_{{\bf n}}$ 
of the special unitary group $\SU(d)$ \cite{GW}.
The tensored space ${\cal H}_A^{\otimes n}$ is decomposed as
(\ref{h4}) by 
\begin{align*}
{\cal H}_A^{\otimes n}=
\bigoplus_{{\bf n}}W_{{\bf n}} , \quad
W_{{\bf n}}:= U_{{\bf n}} \otimes V_{{\bf n}}.
\end{align*}
For the detail, see Weyl \cite{Weyl}, Goodman-Wallch\cite{GW}, 
Iwahori \cite{Iwa}.
The density $\rho_A^{\otimes n}$ is invariant 
w.r.t. the representation of  the $n$-th symmetric group $S_n$ 
on the tensored space ${\cal H}_A^{\otimes n}$.
This type decomposition does not depends on $\rho_A$ and $|\phi
\rangle \langle \phi|$ and depends on the group representation invariance.
But the type of (\ref{h1}) depends on $\rho_A$.
Now, we perform the above invariant entanglement concentration
w.r.t. the subscript ${\bf n}$.
In this case, 
when we get measured value ${\bf n}$,
the final state is the perfect Bell state with the size $\dim V_{{\bf n}}$.
Its probability is $\Tr W_{{\bf n}} \rho^{\otimes n}_A$.
\begin{thm}
The probabilities are evaluated as 
\begin{align*}
\lim_{n \to \infty}\frac{-1}{n}
\log \sum_{{\bf n}}\{
\Tr W_{{\bf n}}\rho_A^{\otimes n}|
\dim V_{{\bf n}}\le 2^{n R}\} 
&= \sup \{ D({\bf q}\| {\bf p})| H( {\bf q}) \le R\}\\
&=
\sup_{s \ge 1}
\frac{(1-s)R- \psi(s) }{s}
\hbox{ if } R \le H(\rho) \\
\lim_{n \to \infty}\frac{-1}{n}
\log \sum_{{\bf n}}\{
\Tr W_{{\bf n}}\rho_A^{\otimes n}|
\dim V_{{\bf n}}\ge 2^{n R}\} 
&= \sup \{ D({\bf q}\| {\bf p})| H( {\bf q}) \ge R\}\\
&=
\sup_{0 \,< s \le 1}
\frac{(1-s)R- \psi(s) }{s}
\hbox{ if } R \ge H(\rho),
\end{align*}
where $\psi(s):= \log \Tr \rho^s_A$ and 
the vector ${\bf p}= (p_1, \ldots, p_d)$ is the set of eigenvalues of
$\rho_A$ satisfying $p_1 \ge p_2 \ge \ldots \ge p_d$.
Thus, when $R \,< H(\rho_A)$,
$\sup_{s \ge 1}
\frac{(1-s)R- \psi(s) }{s} \,> 0$.
\end{thm}
This theorem implies that 
this protocol achieve the bound with the probability which goes to $1$.
The above theorem follows from the following lemmas 
proved in Appendix.
\begin{lem}\Label{l3}
There exists a constant number $C$ such that
\begin{align}
\left|\frac{1}{n}\log \dim V_{{\bf n}} - H\left( \frac{{\bf n}}{n}
\right)\right|
\le \frac{2d^2 +d}{2n}\log (n+d) +
\frac{C}{n},\quad \forall {\bf n}.\Label{l10}
\end{align}
\end{lem}
\begin{lem}\Label{l4}
For any state $\rho_A$ on ${\cal H}_A$ and
any set $R \subset 
R_+ := \{ {\bf p}| p_1 \ge p_2 \ge \ldots \ge p_d\ge 0, \sum_{i=1}^d
p_i =1 \} \subset {\bf R}^d$
and any $\epsilon \,> 0$
there exists $N$ such that 
\begin{align}
\lim_{n \to \infty}
\frac{-1}{n} \log \sum_{\frac{{\bf n}}{n} \in R}
\Tr W_{{\bf n}} \rho^{\otimes n}_A
\ge D( \overline{R}\|{\bf p}):= 
\inf_{ {\bf q} \in \overline{R}}
D( {\bf q}\| {\bf p}),
\end{align}
where 
$\overline{R}$ is the closure of $R$.
\end{lem}
\section{Optimal exponent of universal entanglement concentration}\label{s3.1}
We prove that our universal entanglement concentration is optimal
among universal entanglement concentrations which achieving the optimal rate
$n H(\rho_A)$ for any state.
We call a decomposition $C= \{ C(\omega)\}_{\omega}$ by CP maps of a trace preserving CP map
{\it an instrument}.
We discuss only local operations in 
this section.
A sequence $\{ (C_n =\{ C_n(\omega)\}_{\omega}, H_n) \}$ 
pairs of an instrument consisting of local operations on 
$\{ {\cal H}_A^{\otimes n} \otimes {\cal H}_B^{\otimes n}\}$ and
function $H_n:\omega \mapsto H_n(\omega)$
is called {\it an approximately entanglement concentration} of $|\phi \rangle \langle \phi|$
on ${\cal H}_A\otimes {\cal H}_B$ if 
\begin{align}
\sum_{\omega}\Tr C_n(\omega)
(|\phi \rangle \langle \phi|^{\otimes n})
\left\| 
\frac{ C_n(\omega)(|\phi \rangle \langle \phi|^{\otimes n})}
{\Tr C_n(\omega)(|\phi \rangle \langle \phi|^{\otimes n})}-
|\phi_{H_n(\omega)}\rangle \langle \phi_{H_n(\omega)}|
\right\|_{\Tr} \to 0,
\end{align}
where $|\phi_{H_n(\omega)}\rangle \langle \phi_{H_n(\omega)}|$
is the perfect Bell state with the size $H_n(\omega)$. From 
the monotonicity of the infimum of the relative entropy
$D(|\phi \rangle \langle \phi|,\rho)$ among 
non-entanglement states 
$\rho$ on ${\cal H}_A \otimes {\cal H}_B$,
an approximately entanglement concentration $\{(C_n,H_n)\}$ of 
$|\phi \rangle \langle \phi|$ satisfies that
\begin{align}
\lim_{n \to \infty} \sum_{\omega}
\Tr C_n(\omega)(|\phi \rangle \langle \phi|^{\otimes n})
\frac{H_n(\omega)}{n}
\le H(\rho_A) .\label{h20}
\end{align}
A sequence $\{ (C_n =\{ C_n(\omega)\}_{\omega}, H_n) \}$ 
is called an {\it approximately universal entanglement concentration} of a state
family ${\cal S}:= \{\theta \in \Theta |~|\phi_{\theta} \rangle \langle \phi_{\theta}|\}$
on ${\cal H}_A\otimes {\cal H}_B$ if 
it is an approximately entanglement concentration of any state 
$|\phi_{\theta} \rangle \langle \phi_{\theta}|$ and 
satisfies that
\begin{align}
\lim_{n \to \infty} 
\sum_{\omega}
\left\{\Tr C_n(\omega)(|\phi_{\theta}  \rangle \langle \phi_{\theta}|^{\otimes n})
|
H_n(\omega) \ge n( H(\rho_{A,\theta})- \epsilon)
\right\} = 1, \quad
\forall \epsilon \,> 0, \forall  \theta \in \Theta, \label{h21}
\end{align}
where $\rho_{A,\theta}:= \Tr_B |\phi_{\theta}  \rangle \langle
\phi_{\theta}|$. From (\ref{h20}) and (\ref{h21}),
the equation
\begin{align}
\sum_{\omega}
\left\{\Tr C_n(\omega)(|\phi_{\theta}  \rangle \langle \phi_{\theta}|^{\otimes n})
\left|
\left|\frac{H_n(\omega)}{n}- H(\rho_{A,\theta})\right| \,> \epsilon \right.
\right\} = 1, \quad
\forall \epsilon \,> 0, \forall \theta \in \Theta. \label{h22}
\end{align}
Thus, we can regard the function $\frac{H_n(\omega)}{n}$ as
a consistent estimator of the parameter $ H(\rho_{A,\theta})$
on the state family $\{ \theta \in \Theta|\rho_{A,\theta} \}$.
Therefore, 
we have the following theorem.
\begin{thm}
An approximately universal entanglement concentration 
$\{ ( C_n,H_n)\}$ of a state
family ${\cal S}:= \{\theta \in \Theta |~|\phi_{\theta} \rangle \langle \phi_{\theta}|\}$
on ${\cal H}_A\otimes {\cal H}_B$
satisfies that
\begin{align}
\limsup_{n \to \infty}\frac{-1}{n}\log \sum_{\frac{H_n(\omega)}{n}\in R}
\Tr C_n(\omega)(|\phi_{\theta_0} \rangle \langle \phi_{\theta_0}|^{\otimes n})
\le \inf_{H(\rho_{A,\theta})\in R}
D(\rho_{A,\theta}\|\rho_{A,\theta_0})
\end{align}
for any open set $R \in \real$ and any state 
$|\phi_{\theta_0} \rangle \langle \phi_{\theta_0}|$.
\end{thm} From this theorem, 
we can see that
our universal entanglement concentration
achieves the optimal failure exponent
for the state family of all pure states
on the total system ${\cal H}_A \otimes {\cal H}_B$.

\begin{pf}
We define two probabilities $p_n:=
\sum_{\frac{H_n(\omega)}{n}\in R}\Tr C_n(\omega)
(|\phi _{\theta_0}\rangle 
\langle \phi_{\theta_0}|^{\otimes n})$ and \par
\noindent$q_n:=\sum_{\frac{H_n(\omega)}{n}\in R}\Tr C_n(\omega)
(|\phi_{\theta}  \rangle \langle \phi_{\theta}|^{\otimes n})$
for any state $\rho_{A,\theta}$ satisfying $H(\rho_{A,\theta})\in R$.
Since we can regard $C_n$ as a POVM on ${\cal H}_A^{\otimes n}$,
using the monotonicity of relative entropy
we have
\begin{align}
n D(\rho_{A,\theta}\| \rho_{A,\theta_0})
\ge q_n \log \frac{q_n}{p_n} + (1-q_n)\log\frac{1-q_n}{1-p_n}.
\end{align}
Since it follows from (\ref{h22}) that $q_n \to 1$,
we have 
\begin{align}
\limsup_{n \to \infty}\frac{-1}{n}\log p_n
\le D(\rho_{A,\theta}\|\rho_A).
\end{align}
We obtain the desired assertion.
\end{pf}
\section{Teleportation}\label{s4}
If we perform $R$-qubits teleportation,
we can make $R$ qubits perfect Bell state by LOCC.
Conversely, if we make $R$ qubits perfect bell state by LOCC,
we are possible to perform $R$-qubits teleportation.
%Therefore, it is enough for teleportation to dicuss to entanglement concentration.
In the above setting,
the bound of the number of qubit of teleportation
is $n H(\rho_A)$.

Next, we discuss 
how many classical bits we need to perform 
$n H(\rho_A)$ qubits quantum teleportation in the above setting.
It is clear that we need $2 n H(\rho)$ bits classical information.
Using our universal entanglement concentration,
we can perform it
with $2 n H(\rho)$ bits classical information. From this point of view,
our universal entanglement concentration is effective for 
the teleportation.
\section{Dense coding}\label{s5}
We formulate the effect of dense coding as follows.
We assume that 
the state on the tensored total system 
${\cal H}_A^{\otimes n} \otimes {\cal H}_B^{\otimes n}$
is written by $n$-tensored state $| \phi \rangle \langle \phi 
|^{\otimes n}$,
where $| \phi \rangle \langle \phi |$ is a pure state on 
the single total system ${\cal H}_A \otimes {\cal H}_B$.
we call the quadruple $\Phi^n=(M_n,N_n,C^{(n)}_{\bullet},X^{(n)})$
{\it a code} for $| \phi \rangle \langle \phi 
|^{\otimes n}$
when it consists of a natural number $M_n$ 
(the size of sent classical information),
a natural number $N_n$ (the size of sending quantum state),
a POVM (decoding) $X^{(n)}= \{ X^{(n)}_i \}_{i=1}^{M_n}$ and 
a mapping (encoding) $C_{\bullet}^{(n)}: \{ 1, \ldots, M_n\} \ni i \to
C_{i}^{(n)}$,
where $C_{i}^{(n)}$ is a CP map from ${\cal S}({\cal H}_A^{\otimes n})$
to ${\cal S}({\bf C}^{N_n})$ and ${\cal S}({\cal H})$
denotes the set of densities on ${\cal H}$.
Therefore, the effect of entanglement 
is characterized by the quantity $\log \frac{M_n}{N_n}$.
For a code $\Phi^{(n)}=(M_n,N_n,C_\bullet^{(n)} ,X^{(n)})$,
the average error probability is represented by 
\begin{align*}
{\rm E}[\Phi^{(n)}]=
\frac{1}{M_n} \sum_i 
\Tr (C_i^{(n)} \otimes I)(| \phi \rangle \langle \phi |^{\otimes n})
( I- X^n_i).
\end{align*}
Thus, we focus the following quantity
\begin{align*}
C(| \phi \rangle \langle \phi|):=
\sup \left \{\liminf_{n \to \infty} \frac{1}{n}\log
\frac{M_n}{N_n}\left|
\begin{array}{l}
\exists \{\Phi^{(n)}=(M_n,N_n,C_i^{(n)} ,X^n)\}  
\hbox{s.t. }{\rm E}[\Phi^{(n)}] \to 0 
\end{array}
\right. \right\}.
\end{align*}
We have the following theorem.
\begin{thm}\Label{t2}
\begin{align*}
C(| \phi \rangle \langle \phi|)= H(\rho_B) = H(\rho_A),
\end{align*}
where $\rho_A:=\Tr_B | \phi \rangle \langle \phi|$ and 
$\rho_B:=\Tr_A | \phi \rangle \langle \phi|$.
\end{thm}
\begin{pf}
Define the following quantities:
\begin{align*}
I(P, \rho_{\bullet} , X)&:= \sum_{i} P_i \sum_j 
\Tr X_i \rho_j \log \frac{\Tr X_i \rho_j}{\Tr X_i \overline{\rho}} \\
I(P, \rho_{\bullet} )&:= \sum_{i} P_i 
D(\rho_j \|\overline{\rho}) =
H(\overline{\rho}) - \sum_{i} P_i H(\rho_i),
\end{align*}
where $\overline{\rho}:= \sum_j P_i \rho_j$ and $D(\rho\|\sigma):=
\Tr \rho(\log \rho- \log\sigma)$.
According to Barenco-Ekert\cite{BE},
there exists the set $\{ U_i \}_i$ of unitaries on ${\cal H}_A$
and the probability $P$ on it
such that 
\begin{align}
I(P,U_{\bullet}\otimes I(|\phi \rangle \langle \phi|))
= H(\rho_B)+ \dim {\cal H}_A. \Label{h7}
\end{align}
Using the quantum channel coding theorem in the pure state case \cite{Ha},
we can prove that there exists a code achieving the bound $H(\rho_B)$.

Conversely, we can prove that there does not exists a code exceeding
the bound $ H(\rho_B)$ as follows.
For any density $\sigma$ the relations
\begin{align*}
\sum_{i} P_i  D (\rho_j \| \sigma) 
&= - \sum_{i} P_i H(\rho_i) - \Tr \overline{\rho}   \log  \sigma \\
&= - \sum_{i} P_i H(\rho_i) + H(\overline{\rho})
+ D(\overline{\rho} \|  \sigma) \\
&\ge  - \sum_{i} P_i H(\rho_i) + H(\overline{\rho})
= I( P,\rho_{\bullet})
\end{align*}
hold. Letting $P^{(n)}_i: = \frac{1}{M_n}$,
from Fano's inequality, we have
\begin{align*}
& -\log 2 + (1- {\rm E}[\Phi^{(n)}] )\log M_n \\
& 
\le I( P^{(n)}, C_{\bullet}^{(n)}\otimes I (| \phi \rangle \langle
\phi|^{\otimes n}), X^{(n)}) \\
&\le I( P^{(n)}, C_{\bullet}^{(n)}\otimes I (| \phi \rangle \langle
\phi|^{\otimes n}))\\
& \le  \sum_{i} P_i^{(n)}
D\left(\left. C_{\bullet}^{(n)}\otimes I (| \phi \rangle \langle
\phi|^{\otimes n})\right\| \frac{1}{N_n}I \otimes \rho_B^{\otimes n}\right) \\
&= - \sum_{i} P_i^{(n)}
H(C_{\bullet}^{(n)}\otimes I (| \phi \rangle \langle
\phi|^{\otimes n})) + \log N_n + H(\rho_B^{\otimes n}).
\end{align*}
Therefore, it follows that 
\begin{align*}
H(\rho_B)
&\ge \frac{1}{n}H(\rho_B^{\otimes n})
-  \sum_{i} P_i^{(n)}
H(C_{\bullet}\otimes I (| \phi \rangle \langle
\phi|^{\otimes n})) \\
&\ge \frac{\log M_n - \log N_n}{n}
- \frac{1}{n} \left( \log 2 + {\rm E}[\Phi^{(n)}] \log M_n \right).
\end{align*}
Since $ {\rm E}[\Phi^{(n)}] \to 0$,
we have the converse inequality.
\end{pf}

Using our universal entanglement concentration,
we make the perfect Bell state with the size $\dim V_{{\bf n}}$.
With the probability $1$,
we can send classical information with the size $\dim V_{{\bf n}}^2$
by sending the quantum state with the size $\dim V_{{\bf n}}$.
In this case $M_n= \dim V_{{\bf n}}^2 $ and $N_n= \dim V_{{\bf n}}$.
If $R \le H(\rho_A)$, the probability 
of the relation $\frac{M_n}{N_n} = \dim V_{{\bf n}} \le 2^{n R}$
goes to $0$ with the exponent $\sup_{s \ge 1}\frac{(1-s)R- \psi(s) }{s}$.
This is another proof of the direct part of Theorem \ref{t2}.

Next, we compare its exponent with another protocol.
The Burnashev-Holevo \cite{BH} random coding exponent of the pair
$(\{U_i \}_i,P)$ satisfying (\ref{h7})
is $\sup_{2 \ge s \ge 1}(1-s)R- \psi(s) $,
which is better than 
$\sup_{s \ge 1}\frac{(1-s)R- \psi(s) }{s}$
when $R \,< H(\rho_A)$ is large enough.
\appendix
\section{Proof of Lemma \protect{\ref{l3}}}
According Weyl \cite{Weyl}, Iwahori \cite{Iwa},
the dimension of $V_{{\bf n}}$ is written by 
\begin{align*}
\dim V_{{\bf n}}=\frac{n !}{(n_1+d-1)! (n_2+d-2)! \ldots n_d!}
\prod_{j \,> i}(n_i-n_j-i+j).
\end{align*}
Then the equation 
\begin{align*}
&\frac{1}{n}\log \dim V_{{\bf n}}\\
&= \sum_{i=1}^d
-\frac{n_i}{n}\log \frac{n_i}{n}
+\frac{1}{n}
\sum_{i=1}^d \log \frac{n_i!}{(n_i+d-i)!}
+\frac{1}{2n}\log \frac{n_1 n_2 \ldots n_d}{n} \\
&+\frac{1}{n} \sum_{j \,> i} \log (n_i-n_j-i+j)
+\frac{1}{n}( \delta_n - \sum_{i-1}^d \delta_{n_i} )
\end{align*}
holds, where
$\delta_n$ is defined as
$ n! = e^{\delta_n}n^{n+\frac{1}{2}}e^{-n}$
and converges to the constant $\frac{1}{2}\log 2 \pi$.
Then, we choose the constant $C$ as
$C: = d \sup_n \delta_n$.
Since 
\begin{align*}
&\left |\frac{1}{n}
\left( \sum_{i=1}^d \log \frac{n_i!}{(n_i+d-i)!}
+\frac{1}{2} \log \frac{n_1! n_2! \ldots n_d}{n}
+\sum_{j \,> i} \log (n_i-n_j-i+j)\right)\right|\\
&\le  
\frac{1}{n}\log (n+d)
(\frac{d(d-1)}{2}+\frac{d}{2}+d(d-1))=
\frac{3d^2 -2d}{2n}\log (n+d),
\end{align*}
the inequality (\ref{l10}) holds.
\section{Proof of Lemma \protect{\ref{l4}}}
Define the vectors ${\bf v}^{l}, g {\bf v}$ and ${\bf d}$ by
\begin{align*}
{\bf v}^{l} &: = ( v_1^l , v_2^l , \ldots, v_d^l) \\
g {\bf v}&:= ( v_{g(1)}, v_{g(2)}, \ldots, v_{g(d)}) \\
{\bf d}& := (d-1,d-2, \ldots, 0).
\end{align*}
for any $g \in S_d$.
According to Weyl \cite{Weyl}, Iwahori \cite{Iwa},
the probability $\Tr W_{{\bf n}} \rho_{A}^{\otimes n}$
is written by 
\begin{align*}
\Tr W_{{\bf n}} \rho_{A}^{\otimes n}
=
\dim V_{{\bf n}}
\frac{\det ( {\bf p}^{n_1+d-1}, {\bf p}^{n_2+d-2},\ldots,
{\bf p}^{n_d})}
{\prod_{i \,> j}(p_i - p_j)}.
\end{align*}
According to Weyl \cite{Weyl}, Iwahori \cite{Iwa},
the dimension of $\dim V_{{\bf n}}$
has another form as
\begin{align*}
\dim V_{{\bf n}}= \sum_{g \in S_d}
\sgn (g) C({{\bf n}}+ {\bf d}- g{\bf d}) ,
\end{align*}
where $C_{{\bf n}}$ is defined as
\begin{align*}
C({\bf n}):= \left\{
\begin{array}{ll}
\frac{n!}{n_1! n_2! \ldots n_d !} &\hbox{ if  }
n_i \ge 0, \sum_{i=1}^d n_i =1 \\
0 & \hbox{ otherwise}. 
\end{array}
\right.
\end{align*}
Using the above formula,
we can calculate the probability as
\begin{align*}
\Tr W_{{\bf n}} \rho_{A}^{\otimes n}
&=
\frac{1}{\prod_{i \,> j}(p_i - p_j)}
\sum_{g,g' \in S_d}
\sgn(gg')\prod_{i=1}^d p_i^{n_{g(i)}+d- g(i)}
C({\bf n}+ {\bf d} - g {\bf d}) \\
&= \frac{1}{\prod_{i \,> j}(p_i - p_j)}
\sum_{g,g' \in S_d}
\sgn(gg')\prod_{i=1}^d p_i^{d-g g'(i)}
{\rm Mul}( {\bf p}, g {\bf n}- g g'{\bf d} + g {\bf d})
\end{align*}
where 
we denote the Multinomial distribution of ${\bf p}$ by
${\rm Mul}( {\bf p}, \bullet )$.
For any $\epsilon_1 \,> 0$, there exists 
an integer $N$ such that 
\begin{align*}
\frac{g {\bf n} - g g' {\bf d} + g {\bf d}}{n}
\in U((R_+^c))_{\epsilon_1}.
\end{align*}
$U(R)_{\epsilon_1}:= \cup_{{\bf q}\in R}U_{{\bf q},\epsilon_1}$
and $U_{{\bf q},\epsilon_1}$ is $\epsilon_1$-neighborhood of ${\bf q}$.
It follows from Sanov's theorem that 
for any $\epsilon_2 \,> 0$ there exists
$N$ such that
the inequalities
\begin{align*}
\sum_{\frac{{\bf n}}{n}\in R_+}
{\rm Mul}( {\bf p}, 
g {\bf n}- g g'{\bf d} + g {\bf d})
\le
\sum_{\frac{{\bf n}}{n}\in U((R_+^c))_{\epsilon_1} }
{\rm Mul}( {\bf p}, {\bf n})
\le e^{-n D( U(U((R_+^c)_{\epsilon_1})_{\epsilon_2}\| {\bf p})}
\end{align*}
hold for any $g'\in S_d$, any non-identical element $g \in S_d$
and any $n\ge N$.
For any $\epsilon_3 \,> 0$, there exists an integer $N$
such that
\begin{align*}
\frac{{\bf n}-  g'{\bf d} + {\bf d}}{n}
 \in U(R)_{\epsilon_3}, \forall {\bf n} \in R, \quad \forall n \ge N.
\end{align*} From Sanov's Theorem, for any $\epsilon_4 \,> 0$ there exists
$N$ such that 
\begin{align*}
\sum_{\frac{{\bf n}}{n}\in R}
{\rm Mul}( {\bf p}, 
{\bf n}-  g'{\bf d} + {\bf d})
\le e^{-n D(U(U(R)_{\epsilon_3})_{\epsilon_4}\| {\bf p})}, \quad \forall
n \ge N.
\end{align*}
Letting $D({\bf P}):=
\frac{1}{\prod_{i \,> j}(p_i - p_j)}
\sum_{g,g' \in S_d}
\sgn(gg')\prod_{i=1}^d p_i^{d-g g'(i)}$,
we have
\begin{align*}
& \sum_{\frac{{\bf n}}{n}\in R}
\Tr W_{{\bf n}} \rho_{A}^{\otimes n} \\
& \le
\frac{1}{\prod_{i \,> j}(p_i - p_j)}
\sum_{g' \in S_d}
\sgn(g')\prod_{i=1}^d p_i^{d- g'(i)}
\sum_{\frac{{\bf n}}{n}\in R}
{\rm Mul}( {\bf p}, 
{\bf n}-  g'{\bf d} + {\bf d}) \\
&+
\frac{1}{\prod_{i \,> j}(p_i - p_j)}
\sum_{g'\in S_d}\sum_{\id \neq g \in S_d}
\sgn(gg')\prod_{i=1}^d p_i^{d-g g'(i)}
\sum_{\frac{{\bf n}}{n}\in R}
{\rm Mul}( {\bf p}, 
g {\bf n}- g g'{\bf d} + g {\bf d}) \\
&\le D({\bf P})\left( e^{-n D((U(U(R)_{\epsilon_3})_{\epsilon_4}\| {\bf p})}
+e^{-n D(  U(U(R_+^c)_{\epsilon_1})_{\epsilon_2}\| {\bf p})}\right) .
\end{align*}
Therefore 
\begin{align*}
\lim_{n \to \infty}
\frac{1}{n} \log \sum_{\frac{{\bf n}}{n}\in R}
\Tr W_{{\bf n}} \rho_{A}^{\otimes n} 
\le - \min \{  D(  U(U(R_+^c)_{\epsilon_1})_{\epsilon_2}\| {\bf p}),
 D((U(U(R)_{\epsilon_3})_{\epsilon_4}\| {\bf p})\}.
\end{align*}
From the  arbitrarity of $\epsilon_1, \epsilon_2, \epsilon_3
\epsilon_4 \,>0$, we have
\begin{align*}
\lim_{n \to \infty}
\frac{-1}{n} \log \sum_{\frac{{\bf n}}{n}\in R}
\Tr W_{{\bf n}} \rho_{A}^{\otimes n} 
\ge  D( \overline{R}\| {\bf p}),
\end{align*}
where $\overline{R}$ is the closure of $R$.

\end{document}